Exploration of oxide-based diluted magnetic semiconductors toward transparent spintronics

T. Fukumura[a,*], Y. Yamada[a], H. Toyosaki[a], T. Hasegawa[b,d], H. Koinuma[c,d], M. Kawasaki[a,d]

[a] *Institute for Materials Research, Tohoku University, Sendai 980-8577, Japan*

[b] *Frontier Collaborative Research Center, Tokyo Institute of Technology, Yokohama 226-8503, Japan*

[c] *Materials and Structures Laboratory, Tokyo Institute of Technology, Yokohama 226-8503, Japan*

[d] *Combinatorial Materials Exploration and Technology (COMET), Tsukuba 305-0044, Japan*

A review is given for the recent progress of research in the field of oxide-based diluted magnetic semiconductor (DMS), which was triggered by combinatorial discovery of transparent ferromagnet. The possible advantages of oxide semiconductor as a host of DMS are described in comparison with conventional compound semiconductors. Limits and problems for identifying novel ferromagnetic DMS are described in view of recent reports in this field. Several characterization techniques are proposed in order to eliminate unidentified ferromagnetism of oxide-based DMS (UFO). Perspectives and possible devices are also given.

PACS: 75.50Pp, 75.70-i, 75.30Hx, 81.15Fg

Keywords: magnetic semiconductor, oxide semiconductor, spintronics, ZnO, $TiO_2$, unidentified ferromagnetic oxide, ferromagnetism, epitaxial thin film

[*]Corresponding Author.

Tomoteru Fukumura

Institute for Materials Research

Tohoku University

Sendai 980-8577, Japan.

Tel. +81-22-215-2084; Fax. +81-22-215-2086.

Electronic mail: fukumura@imr.tohoku.ac.jp

## 1. Introduction

Diluted magnetic semiconductor (DMS) is expected to play an important role in interdisciplinary materials science and future spintronics because charge and spin degrees of freedom are accommodated into single matter and their interplay is expected to explore novel physics and new devices. Among them, Mn-doped II-VI [1] and III-V [2] compound semiconductors have been extensively studied. The former include a variety of compounds consisting of various combinations of II-group cations (Zn, Cd, and Hg) and VI-group anions (S, Se, and Te), some of which have been applied to magneto-optical devices [3]. On the other hand, the latter can be ferromagnetic materials, where one can control the ferromagnetic properties with electric field or light and one can inject spin-polarized carriers into heteroepitaxial semiconductor device [4-6]. However, the ferromagnetic Curie temperature $T_C$ has been much lower than room temperature, *e.g.* $T_C$ ~110 K in (Ga,Mn)As, hence new DMS having $T_C$ beyond room temperature is desired for future devices.

Generally, oxide semiconductors have wide bandgap, *i.e.* transparent for visible light, and can be doped heavily with *n*-type carrier. This feature serves an important role as transparent conductor that is used for various applications [7]. From the viewpoint of DMS, this feature can be promising for strong ferromagnetic exchange coupling between localized spins due to carrier induced ferromagnetism such as Ruderman-Kittel-Kasuya-Yosida interaction and double exchange interaction when localized spin is introduced in the oxide semiconductor. This situation prompted us to make oxide-based DMS based on a representative oxide semiconductor ZnO [8]. However, our intensive research on combinatorial thin film preparation revealed that none of transition metal (TM) doped ZnO was ferromagnetic down to ~3 K [9,10], whereas band calculation studies suggested the possible ferromagnetism in case of *p*-type ZnO host [11-13]. On the other hand, we have found that Co-doped $TiO_2$ is ferromagnetic for both anatase and rutile phases above room temperature [14,15]. Our efforts to explore transparent ferromagnet triggered intensive research in this field by many other groups employing traditional, *i.e.* not combinatorial, thin film growth techniques. Some of them reveal important characteristics of this class of materials, whereas some claim ferromagnetism through rather crude characterization of their specimens. Here, we overview a short history of oxide-based DMS to reexamine various characterization techniques which have been frequently exploited to claim ferromagnetic oxide-based DMS and to propose much better analysis to make the claims clear.

## 2. DMSs based on various oxide semiconductors

There are many non-oxide semiconductors. Compared to them, the advantages of oxide semiconductors are: (1) wide bandgap suited for applications with short wavelength light, (2) transparency and dyeability with pigments, (3) high *n*-type carrier concentration, (4) capability to be grown at low temperature even on plastic substrate, (5) ecological safety and durability, (6) low cost, etc. In addition, large electronegativity of oxygen is expected to produce strong *p*-*d* exchange coupling between band carriers and localized spins [16]. Such advantages make oxide semiconductors attractive. Actually, many studies on oxide-based DMS have been reported as summarized in Table 1, where most of the researches employed ZnO and $TiO_2$ as host semiconductors.

### a. ZnO

After the first study of Mn-doped ZnO [8], many studies on ZnO doped with various kinds of TMs have been reported. However, the claims concerning the existence of ferromagnetism have diverged. Several studies claim nonferromagnetic behavior of ZnO doped with TM [8-10,17,18], whereas the other groups claim ferromagnetic behavior of the same compounds [19-27]. The reported values of $T_C$ scatter from 30 K to 550 K.

### b. $TiO_2$

The anatase and rutile phases of Co-doped $TiO_2$ were reported to be ferromagnetic as the first trial of these compounds [14,15]. Several subsequent reports agree with the results [28-31], whereas the other studies report that the precipitation of Co metal is the origin of ferromagnetic signal [32-34]. There have been few reports on the definite value of $T_C$ because $T_C$ is too high to be measured by conventional tools such as magnetometer employing superconducting quantum interference devices (SQUID). Doping of TM elements except for Co has been scarcely reported.

### c. $SnO_2$

Mn-doped $SnO_2$ shows a large magnetoresistance and the magnetization behavior is paramagnetic down to 5 K [35]. There have been very few reports on this compound so far.

## 3. Experimental evidences for ferromagnetism in oxide-based DMS

### a. Extrinsic effects

There has been controversy on the magnetic properties of oxide-based DMS as stated above. Most probable reason is originated from insufficient characterization of the samples. Deduction of magnetic property only from magnetization measurements without careful examination of possible extrinsic effects, such as ferromagnetic precipitate and impurity phase, often misleads us into creating unidentified ferromagnetic oxides (UFO). In order to evaluate the magnetism correctly, various characterization techniques have to be employed. Table 2 shows a list of characterization techniques to eliminate UFO. X-ray diffraction measurement is indispensable to detect impurity phases, but the sensitivity may not be good enough to identify a small amount of precipitation in the sample. Such precipitation can be detected with other techniques. Scanning electron microscope and reflection high energy electron diffraction can examine the precipitation at the sample surface and transmission electron microscope can even identify the precipitation in the sample on the scale of nanometer. Electron probe microanalysis and secondary ion mass spectrometry examine uniformity of dopant distribution along lateral and perpendicular directions of the sample, respectively. Most of the papers in Table 1 have been reported with employing only part of these techniques.

### b. Magnetic properties

Ferromagnetic DMS has various properties, some of which are unique for identification of the ferromagnetism in oxide-based DMS. Table 3 shows various characterization techniques for magnetic properties to identify ferromagnetism to eliminate UFO. Conventional DMS has $T_C$ that depends on carrier concentration and content of TM ion. Hence the magnetization measurements for samples with systematically varied parameters should give systematic variation of $T_C$. Regular pattern of magnetic domain structure on the scale of submicrometer is an evidence of spatially uniform ferromagnetism ruling out the possibility of randomly distributed ferromagnetic precipitation.

Magneto-optical spectroscopy probes the magneto-optical signal as a function of photon energy. Particularly, magnetic circular dichroism (MCD) spectroscopy is useful for thin film study because effect of substrate on the spectrum is negligible contrary to magnetization measurement [36]. The MCD signal is generally enhanced at the absorption edge of host semiconductor of DMS [37-39], as a result of carrier mediated exchange interaction between localized spins. Figure 1(a) shows the absorption spectrum at 300 K and MCD spectrum at 5 K for Co-doped ZnO. The MCD signal appears at the bandgap (3.4 eV) and *d*-*d* transition of $Co^{2+}$ ion (~2 eV), representing proper substitution of Co ion with Zn site. The MCD signal is proportional to applied magnetic field indicating paramagnetic behavior. One of the groups reported ferromagnetic Co-doped ZnO based on solely magnetization measurement [19]. However, MCD spectrum of their sample was large for wide range of photon energy without change in sign. Such behavior is similar to ferromagnetic metal. This feature concludes that the ferromagnetism in their Co-doped ZnO is most likely caused by ferromagnetic precipitation [40]. Figure 1(b) shows absorption and MCD spectra for Co-doped anatase $TiO_2$ at 300 K. The MCD signal is negative around ~2 eV and positive around 3.5-4.5 eV with a peak at 3.57 eV around the bandgap showing ferromagnetic hysteresis for each photon energy. This feature is different from that of Co metal. Thus, the possibility of bulk precipitation of Co metal can be excluded.

As for magnetotransport properties, ferromagnetic DMS such as $Ga_{1-x}Mn_xAs$ shows anomalous Hall effect. On the other hand, Mn-doped ZnO with high carrier concentration shows hysteretic magnetoresistance below 0.2 K implying ferromagnetic ordering, where anomalous Hall effect is not seen [41]. The anomalous Hall effect or the hysteretic magnetoresistance seems to be an evidence of ferromagnetism, whereas the appearance of anomalous Hall effect or hysteretic magnetoresistance has not been reported for Co-doped $TiO_2$. The tunneling measurement such as tunneling magnetoresistance in ferromagnetic tunneling junction [42] and differential tunneling current spectroscopy of tunneling junction with superconducting counter-electrode [43] can be used to evaluate spin polarization of the sample that has to be different from the possible ferromagnetic precipitation.

It is noted that the above magnetic properties are not necessarily required because those are mainly deduced from the studies of typical compound semiconductors having zinc blend or wurtzite crystal structures, where the physics has been understood deeply. The oxide-based DMSs have various crystal structures, hence they have various energy band structures. Therefore, it might not be strange that the magnetic properties of oxide-based DMS are different from those of conventional DMS.

## 4. Conclusion and remarks

The research field of oxide-based DMS is rapidly growing up and many kinds of compounds have been reported. For further progress in this field, minimum requirements are the fabrication of high quality sample, its detail characterization, and the elucidation of origin of observed ferromagnetism as described in this paper, otherwise confusion would emerge to prohibit the advance. After the identification of *true* ferromagnetic oxide-based DMS with high $T_C$, there are various applications. The most attractive feature is the wide bandgap and transparency. It is compatible with blue or ultraviolet laser recently developed, thus magneto-optical application with short wavelength is promising. The oxide semiconductor can be grown on glass or plastic substrate, hence drastic reduction of fabrication cost is possible. Several compounds might have biocompatibility so that an application to biotechnology could be expected. Finally, there are a few *p*-type oxide semiconductors and no report on *p*-type ferromagnetic oxide-based DMS. Materials design and synthesis of such compounds would make transparent spintronics more attractive and feasible.

## Acknowledgements

This work was partly supported by the Ministry of Education, Culture, Sports, Science and Technology, Grant-in-Aid for Creative Scientific Research (14GS0204), the Inamori Foundation, and the Murata Science Foundation.

## References

[1] J. K. Furdyna, J. Appl. Phys. 64 (1988) R29.

[2] H. Ohno, Science 281 (1998) 951.

[3] K. Onodera, T. Masumoto, M. Kimura, Electron. Lett. 30 (1994) 1954.

[4] S. Koshihara, A. Oiwa, M. Hirasawa, S. Katsumoto, Y. Iye, C. Urano, H. Takagi, H. Munekata, Phys. Rev. Lett. 78 (1997) 4617.

[5] H. Ohno, D. Chiba, F. Matsukura, T. Omiya, E. Abe, T. Dieti, Y. Ohno, K.Ohtani, Nature 408 (2000) 944.

[6] Y. Ohno, D. K. Young, B. Beschoten, F. Matsukara, H. Ohno, D. D. Awschalom, Nature 402 (1999) 790.

[7] T. Minami, MRS Bull. 25 (2000) 38.

[8] T. Fukumura, Z. Jin, A. Ohtomo, H. Koinuma, M. Kawasaki, Appl. Phys. Lett. 75 (1999) 3366.

[9] T. Fukumura, Z. Jin, M. Kawasaki, T. Shono, T. Hasegawa, S. Koshihara, H. Koinuma, Appl. Phys. Lett. 78 (2001) 958.

[10] Z. Jin, T. Fukumura, M. Kawasaki, K. Ando, H. Saito, T. Sekiguchi, Y. Z. Yoo, M. Murakami, Y. Matsumoto, T. Hasegawa, H. Koinuma, Appl. Phys. Lett. 78 (2001) 3824.

[11] T. Dietl, H. Ohno, F. Matsukura, J. Cibert, D. Ferrand, Science 287 (2000) 1019.

[12] K. Sato, H. Katayama-Yoshida, Jpn J. Appl. Phys. 39 (2000) L555.

[13] K. Sato, H. Katayama-Yoshida, Jpn J. Appl. Phys. 40 (2001) L651.

[14] Y. Matsumoto, M. Murakami, T. Shono, T. Hasegawa, T. Fukumura, M. Kawasaki, P. Ahmet, T. Chikyow, S. Koshihara, H. Koinuma, Science 291 (2001) 854.

[15] Y. Matsumoto, R. Takahashi, M. Murakami, T. Koida, X.-J. Fan, T. Hasegawa, T. Fukumura, M. Kawasaki, S. Koshihara, H. Koinuma, Jpn J. Appl. Phys. 40 (2001) L1204.

[16] T. Mizokawa, T. Nambu, A. Fujimori, T. Fukumura, M. Kawasaki, Phys. Rev. B65 (2002) 085209.

[17] J.-H. Kim, H. Kim, D. Kim, Y.-E. Ihm, W.-K. Choo, J. Appl. Phys. 92 (2002) 6066.

[18] A. Tiwari, C. Jin, A. Kvit, D. Kumar, J. F. Muth, J. Narayan, Solid State. Commun. 121

(2002) 371.

[19] K. Ueda, H. Tabata, T. Kawai, Appl. Phys. Lett. 79 (2001) 988.

[20] T. Wakano, N. Fujimura, Y. Morinaga, N. Abe, A. Ashida, T. Ito, Physica E10 (2001) 260.

[21] H. Saeki, H. Tabata, T. Kawai, Solid State Commun. 120 (2001) 439.

[22] Y.-M. Cho, W.-K. Choo, H. Kim, D. Kim, Y.-E. Ihm, Appl. Phys. Lett. 80 (2002) 3358.

[23] H.-J. Lee, S.-Y. Jeong, C. R. Cho, C.-H. Park, Appl. Phys. Lett. 81 (2002) 4020.

[24] S. W. Jung, S.-J. An, G..-C. Yi, C. U. Jung, S.-I. Lee, S. Cho, Appl. Phys. Lett. 80 (2002) 4561.

[25] S-J. Han, J. W. Song, C.-H. Yang, S. H. Park, J.-H. Park, Y. H. Jeong, K. W. Rhie, Appl. Phys. Lett. 81 (2002) 4212.

[26] S.-G. Yang, A.-B. Pakhomov, S.-T. Hung, C.-Y. Wong, IEEE Trans. Mag. 38 (2002) 2877.

[27] D. P. Norton, S. J. Pearton, A. F. Hebard, N. Theodoropoulou, L. A. Boatner, R. G. Wilson, Appl. Phys. Lett. 82 (2003) 239.

[28] S. A. Chambers, S. Thevuthasan, R. F. C. Farrow, R. F. Marks, J. U. Thiele, L. Folks, M. G. Samant, A. J. Kellock, N. Ruzycki, D. L. Ederer, U. Diebold, Appl. Phys. Lett. 79 (2001) 3467.

[29] W. K. Park, R. J. O.-Hertogs, J. S. Moodera, A. Punnoose, M. S. Seehra, J. Appl. Phys. 91 (2002) 8093.

[30] N.-J. Seong, S.-G. Yoon, C.-R. Cho, Appl. Phys. Lett. 81 (2002) 4209.

[31] S.R. Shinde, S.B. Ogale, S. Das Sarma, J. R. Simpson, H. D. Drew, S.E. Lofland, C. Lanci, J. P. Buban, N. D. Browning, V.N. Kulkarni, J. Higgins, R.P. Sharma, R.L. Greene, T. Venkatesan, cond-mat/0203576.

[32] D. H. Kim, J. S. Yang, K. W. Lee, S. D. Bu, T. W. Noh, S.-J. Oh, Y.-W. Kim, J.-S. Chung, H. Tanaka, H. Y. Lee, Kawai, Appl. Phys. Lett. 81 (2002) 2421.

[33] P. A. Stampe, R. J. Kennedy, Y. Xin, J. S. Parker. J. Appl. Phys. 92 (2002) 7114.

[34] J.-Y. Kim, J.-H. Park, B.-G. Park, H.-J. Noh, S.-J. Oh, J. S.Yang, D.-H. Kim, S. D. Bu, T.-W. Noh, H.-J. Lin, H.-H. Hsieh, C.T. Chen, Phys. Rev. Lett. 90 (2003) 17401.

[35] H. Kimura, T. Fukumura, H. Koinuma, M. Kawasaki, Appl. Phys. Lett. 80 (2001) 94.

[36] Most of the experimental evidences to claim ferromagnetism in oxide-based DMS have

been magnetization measurements. Magnetization vs. magnetic field curve shows hysteresis and saturation behaviors. The magnetic moment measured by magnetometer is in proportion to the sample volume. In case of the measurement of thin film on substrate, the measured signal includes the magnetic moment not only from the film but also from the substrate owing to the much larger volume of substrate than that of film. Careful analysis of magnetization measurements must be carried out.

[37] K. Ando, T. Hayashi, M. Tanaka, A. Twardowski, J. Appl. Phys. 83 (1998) 6548.

[38] K.Ando, H.Saito, Z. Jin, T. Fukumura, M. Kawasaki, Y. Matsumoto, H. Koinuma, J. Appl. Phys. 89 (2001) 7284.

[39] K. Ando, H. Saito, Zhengwu Jin, T. Fukumura, M. Kawasaki, Y. Matsumoto, H. Koinuma, Appl. Phys. Lett. 78 (2001) 2700.

[40] K. Ando, cond-mat/0208010.

[41] T. Andrearczyk, J. Jaroszynski, M. Sawicki, LeVan Khoi, T. Dietl, D. Ferrand, C. Bourgognon, J. Cibert, S. Tatarenko, T. Fukumura, Z. Jin, H. Koinuma, M. Kawasaki, Proc. of the 25th Int.Conf. on the Physics of Semiconductors, Parts I and II, 234-235, 2001.

[42] S. Maekawa, U. Gafvert, IEEE Trans. Magn. 18 (1982) 707.

[43] P. Meservy, P. M. Tedrow, Phys. Rep. 238 (1994) 173.

Figure captions

Fig. 1. (a) Absorption (dotted line) and MCD (solid line) spectra for Co-doped ZnO (001) epitaxial thin films on sapphire (001) substrates. Co content and measurement temperature are 5 mol% at 300 K and 1 mol% at 5 K, respectively. Both spectra around *d*-*d* transition of $Co^{2+}$ ion are magnified. (b) Absorption (dotted line) and MCD (solid line) spectra for Co-doped anatase $TiO_2$ (001) epitaxial thin film on $LaSrAlO_4$ (001) substrates. Co content and measurement temperature are 10 mol% at 300 K. Applied magnetic field is $10^4$ Oe for MCD measurement.

Table 1

List of oxide-based DMSs recently reported.

| Compound | TM cont. | Substrate | Fabrication method | Growth temp. [°C] | Oxygen pressure [Torr] | Post-annealing | *Tc* [K] | Notes | Ref. |
|---|---|---|---|---|---|---|---|---|---|
| ZnO:Mn | < 0.35 | *c*-sapphire | PLD | 600 | $5\times10^{-5}$ | | N/A | | [8] |
| ZnO:Mn | 0.36 | *c*-sapphire | PLD | 600 | $5\times10^{-5}$ | | N/A | spin-glass | [9] |
| $Zn_{1-x}TM_xO$ | | *c*-sapphire | PLD | 500-600 | $1\times10^{-9}$-$10^{-6}$ | | N/A | | [10] |
| ZnO:Co | 0.02-0.5 | *c*-sapphire | PLD | 300-700 | $1\times10^{-6}$-$10^{-1}$ | | N/A | spin-glass | [17] |
| ZnO:Mn | 0.01-0.36 | *c*-sapphire | PLD | 610 | $5\times10^{-5}$ | | | paramagnetic | [18] |
| ZnO:(Co,Mn,Cr, or Ni) | 0.05-0.25 | *r*-sapphire | PLD | 350-600 | $2$-$4\times10^{-5}$ | | 280-300 | $2\mu_B$/Co | [19] |
| ZnO:Ni | 0.01-0.25 | *c*-sapphire | PLD | 300-700 | $1\times10^{-5}$ | | | superpara- or ferro-magnetic | [20] |
| ZnO:V | 0.05-0.15 | *r*-sapphire | PLD | 300 | $10^{-5}$-$10^{-3}$ | | > 350 | $0.5\mu_B$/V | [21] |
| ZnO: (Co,Fe) | < 0.15 | $SiO_2$/Si | magnetron sputtering | 600 | $2\times10^{-3}$ | 600°C, 10min, $1.0\times10^{-5}$Torr | > 300 | 12-15emu/cm$^3$ | [22] |
| ZnO:Co | 0-0.25 | *c*-sapphire | sol-gel | < 350 | | 700°C, 1min | > 350 | $0.56\mu_B$/Co | [23] |
| ZnO:Mn | 0-0.3 | *c*-sapphire | PLD | | | | > 30-45 | 0.15-0.17$\mu_B$/Mn | [24] |
| ZnO:(Fe,Cu) | 0-0.1 | | solid state reaction | 897 | | | 550 | $0.75\mu_B$/Fe | [25] |
| ZnO:(Co,Al) | 0.04-0.12 | glass | RF sputtering | | $1\times10^{-2}$ in Ar | | > 350 | $0.21\mu_B$/Co | [26] |
| ZnO:(Mn,Sn) | 0-0.3 | | implantation | | | 5 min, 700°C | 250 | | [27] |
| anatase $TiO_2$:Co | < 0.08 | $LaAlO_3$ $SrTiO_3$ | PLD | 677-727 | $1\times10^{-6}$-$10^{-5}$ | | > 400 | $0.32\mu_B$/Co | [14] |
| anatase $TiO_2$:Co | 0.01-0.1 | $SrTiO_3$ | OPA-MBE | 550-750 | $2\times10^{-5}$ | | > 300 | $1.26\mu_B$/Co | [28] |
| anatase $TiO_2$:Co | 0.01-0.1 | Si | sputtering | 250-450 | $5\times10^{-5}$-$3\times10^{-4}$ | | > 400 | $0.94\mu_B$/Co | [29] |
| anatase $TiO_2$:Co | 0.03-0.12 | SiO2/Si | MOCVD | 400-500 | 1 | 700°C, 1h, $1.0\times10^{-6}$ Torr | > R.T. | 20-40emu/cm$^3$ | [30] |
| anatase $TiO_2$:Co | 0-0.15 | $LaAlO_3$ $SrTiO_3$ | PLD | 700 | $1\times10^{-5}$-$10^{-4}$ | | 650-700 | $1.4\mu_B$/Co | [31] |
| anatase $TiO_2$:Co | 0.04 | $SrTiO_3$ | PLD | 600 | $1\times10^{-7}$-$10^{-4}$ | | | Co metal 1.7-2.3$\mu_B$/Co | [32] |
| anatase $TiO_2$:Co | 0.01-0.07 | $LaAlO_3$ $SrTiO_3$ | PLD | 750 | $1\times10^{-3}$ | | | Co cluster $1.7\mu_B$/Co | [33] |
| anatase $TiO_2$:Co | 0.04-0.1 | $LaAlO_3$ | PLD | 650 | $10^{-6}$-$10^{-5}$ | 400°C, 20min, $10^{-6}$ Torr | | Co cluster $1.55\mu_B$/Co | [34] |
| rutile $TiO_2$:Co | 0-0.05 | *r*-sapphire | PLD | 700 | $1\times10^{-6}$ | | > 400 | $1\mu_B$/Co | [15] |
| $SnO_2$:(Mn,Sb) | 0.05-0.34 | *r*-sapphire | PLD | 300-600 | $1\times10^{-2}$ | | N/A | paramagnetic | [35] |

Table 2

Characterization techniques to examine the sample quality.

| Method | Purpose |
|---|---|
| X-ray diffraction | Impurity phase |
| Scanning electron microscope<br>Reflection high energy electron diffraction | Surface precipitation |
| Transmission electron microscope | Microscopic precipitation |
| Electron probe microanalysis<br>Secondary ion mass spectrometry | Uniformity of composition |

Table 3

Various material properties and the expected results in case of ferromagnetic DMS.

| Method | Expected result |
| --- | --- |
| Magnetization measurements | $T_C$ depends on carrier concentration and content of transition metal ion |
| Magnetic domain observation | Regular magnetic domain structure |
| Magnetic circular dichroism | Large signal at absorption edge |
| Magnetotransport | Anomalous Hall effect or hysteretic magnetoresistance |
| Tunneling measurements | Spin polarization different from that of precipitation |

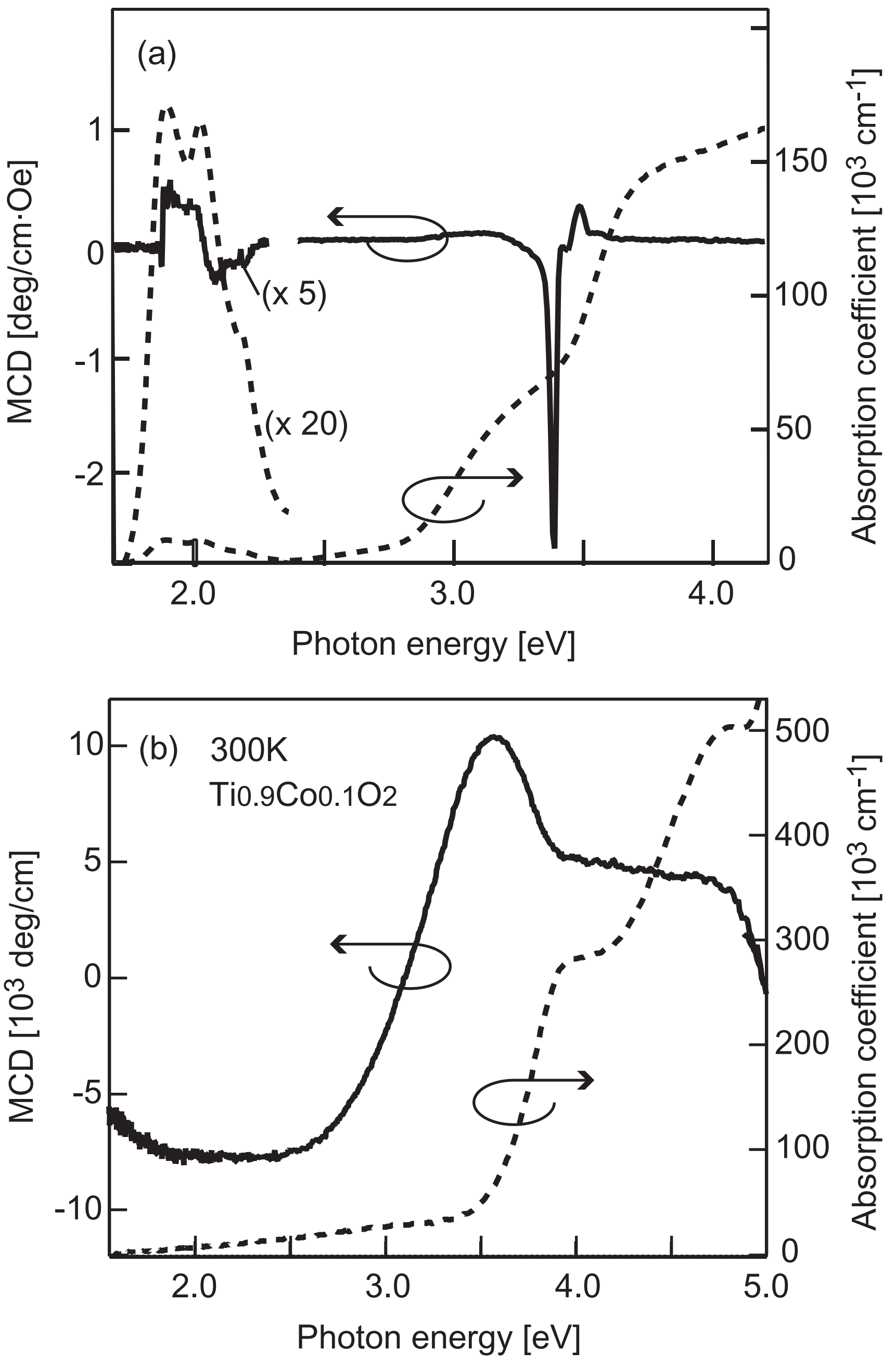

Fig. 1 T. Fukumura et al.